\documentclass[prb,twocolumn,showpacs,preprintnumbers,amsmath,amssymb]{revtex4}

\usepackage{dcolumn}
\usepackage{bm}

\usepackage{graphicx}
\usepackage{graphics}
\usepackage[usenames,dvipsnames]{color}

\newenvironment{frcseries}{\fontfamily{frc}\selectfont}{}

\begin{document}
\title{A Statistical Mechanical Approach to Protein Aggregation}

\author{John S. Schreck}
\email{jss74@drexel.edu}
\author{Jian-Min Yuan}%
\email{yuanjm@drexel.edu}

\affiliation{Department of Physics, Drexel University, Philadelphia, PA 19104}%

\date{\today}

\begin{abstract}             
We develop a theory of aggregation using statistical mechanical methods. An example of a complicated aggregation system with several levels of structures is peptide/protein self-assembly. The problem of protein aggregation is important for the understanding and treatment of neurodegenerative diseases and also for the development of bio-macromolecules as new materials. We write the effective Hamiltonian in terms of interaction energies between protein monomers, protein and solvent, as well as between protein filaments. The grand partition function can be expressed in terms of a Zimm-Bragg-like transfer matrix, which is calculated exactly and all thermodynamic properties can be obtained. We start with a two-state treatment that can be easily generalized to three or more states using a Potts model, for which the exactly solvable feature of the model remains. We focus on $n \times N$ ladder systems, corresponding to the ordered structures observed in some real fibrils. We have obtained results on nucleation processes and phase diagrams, in which a protein property such as the aggregate concentration is expressed as a function of the initial protein concentration and inter-protein or interfacial interaction energies. We have applied our methods to A$\beta$(1-40) and Curli fibrils and obtained results in good agreement with experiments. 

\end{abstract}
\maketitle
\section{Introduction}
In addition to folding into unique native structures of globular proteins, a general property of a protein is its self-assembly into aggregates and fibrils under certain conditions~\cite{dobson}.  Unlike the reversible native structure, formation of solid fibrils could be irreversible and results in an overall stable state of a protein. Aggregates, fibrils, and plaques are often associated with human neurodegenerative diseases, such as Alzheimer's disease, Parkinson's disease, diabetes, and prion-related diseases, to name just a few.  For this purpose, it is important to understand the mechanisms and pathways of the associated aggregation processes.

Due to the high degrees of freedom involved, protein aggregation processes are a difficult problem to study using all-atom molecular dynamics methods.  Simplifying approximations~\cite{zhou} or coarse-graining models are often introduced. Another alternative is to use statistical mechanics and thermodynamics approaches, due to their ability to reduce greatly the number of degrees of freedoms or parameters involved.  Several such statistical mechanical approaches have recently appeared in the literature~\cite{oosawa, ben_shaul, nyrkova, vander, dutch_2001, dutch_2003, dutch_2006, dutch_2008, kunes, nicodemi, lee, zamparo, schmit}.  For example, van Gestel, et al. have developed a simple two-state model for studying helix-coil or sheet-coil transitions in aggregates along with a polymerization transition~\cite{dutch_2001, dutch_2003, dutch_2006, dutch_2008}.  Schmidt, et al.~\cite{schmit} and others~\cite{lee, tiana} focus a well-defined pathway of aggregation including monomer, oligomer, and fibril structures. Zamparo et al.~\cite{zamparo} generalized the WSME model~\cite{wsme1, wsme4} for the studies of protein aggregation that includes helix-sheet transitions. Earlier, Skolnick et al.~\cite{skolnick} and others~\cite{qian, ghosh_dill} used the Zimm-Bragg model for protein folding~\cite{zimm_bragg, lifson_roig, helix_book} to study tertiary interactions between neighboring helical proteins. 

The Zimm-Bragg~\cite{vander, dutch_2001, dutch_2003, dutch_ab, dutch_2006, dutch_2008} or Ising-like models~\cite{zamparo} have been extended and applied to the study of protein aggregation problems, starting from effective Hamiltonians or partition functions.  In one example, van Gestel, et al. assumed a bond linking two proteins can assume coil and helix states~\cite{dutch_2001, dutch_2003} or coil and sheet states~\cite{dutch_2006, dutch_2008}. In Zamparo's models, a protein can take helix or sheet conformations~\cite{zamparo}. In reality, a protein can take all three (or more) conformations~\cite{hong, schreck}, resulting in richer pathways and properties.  Thus, it may be advantageous to introduce a three-state, or more generally, a $q$-state model, where $q$ = 2, 3, 4,$\dots$, an integer.  This can easily be accomplished by using a $q$-state Potts model~\cite{wu}. Another power of the Zimm-Bragg type of approaches is the use of transfer matrices, providing the possibility of obtaining exact or analytic solutions.  The exactly solvable feature can be kept in a Potts model.  

The purpose of the present article is to develop a statistical mechanical theory on protein aggregation based on an effective Hamiltonian, a Potts model, partition functions, and transfer matrices. From this theory, we can obtain thermodynamic and nucleation properties associated with the self-assembly process of proteins. In the next section we describe the system, the aggregation pathways that we investigate, and the aggregate and solution phases. We also describe effective Hamiltonians for a single aggregate and the statistical mechanical methods that are used. In Section 3 we explicitly include solvent (water) interactions and define an effective Hamiltonian for the formation of critical nuclei. We then calculate a few experimentally relevant thermodynamic quantities. In Section 4, we include inter-filament interactions to model full fibrils. In Section 5, our theory is applied to the aggregation of A$\beta$(1-40) and Curli fibril systems, and results are compared to experimental observations. Finally in Section 6, we discuss a helix-sheet-coil aggregation model.

\section{ Systems Studied }
\label{systems}
We consider the protein aggregation pathway from monomers, to dimer, trimers,$\dots$, oligomers,$\dots$, filaments, proto-fibrils and fibrils. In general not all oligomers or aggregates are stable, but monomers, fibrils, and sometimes oligomers are observable on experimental time-scales. Here, we assume that all these species are in kinetic equilibrium and are interested in thermodynamic properties of aggregates. We assume the monomer is an unstructured protein, but in reality it can be collapsed coil~\cite{zhang2000}, which could be taken into consideration in a more detailed model. A filament is a linear chain of interacting, identical proteins that we fix to it a 1D (or quasi-1D) lattice. This is reasonable because aggregates from oligomers to proto-fibrils are still soluble and floating around in solution. Initially, we focus on one of them at a time. The coordinate is along the sequence of the chain and does not necessarily imply the chain is geometrically straight. We consider the chain as a sub-system in a volume solution. Several filaments or proto-fibrils are known to assemble into fibrils, where participating filaments and proto-fibrils are held together by stabilization interactions. In our studies, these structures are put onto strip lattices ($2 \times N$, $3 \times N$, $\dots$, $n \times N$) by which we can model lateral interactions between between $n$ filaments that comprise a single fibril. 

\textcolor{black}{The aggregate phase is the strip (or 1D) lattice that may be occupied by aggregates and any other species, including solvent clusters. This phase is in equilibrium with dissolved proteins in the solvent phase. The chemical potential for protein monomers in the solution can be written~\cite{ferrone, ferrone1, hill} as
\begin{equation}
\label{chemsoln}
\mu_{soln} = \mu_{ST} + \mu_{SR} + R T \ln c
\end{equation}
where the subscript `S' stands for solution, $\mu_{ST}$ and $\mu_{SR}$ are the free energy contributions arising from the translational and rotational motional freedom that monomers possess in solution, respectively, and $c$ is the concentration of monomers in solution. For the chemical potential of the aggregates, $\mu_{agg}$, we assume a crystalline approximation so that $\mu_{agg}$ can be written as~\cite{abe}
\begin{equation}
\label{chemagg}
\mu_{agg}  = \mu_{PC} + \mu_{PV}
\end{equation}
where `P' stands for polymer of proteins. $\mu_{PC}$ and $\mu_{PV}$ are the free energy contributions arising from the contact and interface interactions between proteins in aggregates, and the vibrational motional freedom that proteins in aggregates possess, respectively. Equilibrium between a solution phase and an aggregate phase of protein is then given by
\begin{equation}
\label{equib}
\mu_{agg} = \mu_{soln}.
\end{equation}
With the simple statistical mechanical model presented in the sections to follow, we can relate the chemical potential contribution from the interactions between proteins in aggregates, $\mu_{PC}$, to the experimental concentration of protein in solution via Eq.~(\ref{equib}). We present several different versions of effective Hamiltonians for describing the interactions between proteins in aggregates below.  }

Perhaps the most salient feature of amyloid fibrils is the cross-beta structure~\cite{nelson, luhrs, sawaya, wasmer, petkova}, but conformations such as helix and coil may play roles in the early stages of fibrillization. Our aggregation  model does not start with residue-residue interactions, but with individual protein molecules, which are classified into coil, helix, and sheet proteins, as defined below. In a Zimm-Bragg-like model, order parameters, $\theta$, for the protein are defined as the fractions of that secondary structure in a protein~\cite{helix_book}. When the protein is completely unfolded/folded, $\theta$=0,1, respectively. In our model, a `sheet' protein is one which is dominated by sheet or hairpin structures where on average $\theta_{sheet}>\theta_{helix}$ and $\theta_{sheet}>\theta_{coil}$, which means that the majority of the residues are involved in the formation of sheet structure. A `helix' protein is defined similarly; on average $\theta_{helix} >\theta_{sheet}$ and $\theta_{helix} >\theta_{coil}$ and the protein is majority helical. The random coil is short of secondary structures. To reduce the number of parameters needed to describe protein aggregation, we don't specify conformations other than helix, sheet, or coil. Generally, any number of stable conformations could be included in a model description, and thus instead of using an Ising-like model, we express our Hamiltonian in terms of a Potts model with $q$ states with $q=1,2,3,\dots$ 

A simple effective Hamiltonian for the interactions between $N$ proteins that compose a single filament on a 1D lattice, where the protein could be in a helical, sheet, or coil conformation, can be written in terms of a three-state Potts model as
\begin{eqnarray}
\label{potts}
-\beta \mathcal{H}_{fil} &=& P_{1} \sum_{i=1}^{N-1} \delta( t_{i}, 1 ) + P_{2} \sum_{i=1}^{N-1} \delta( t_{i}, 2 ) \\
&-& \sum_{i=1}^{N-1} R(t_{i}, t_{i+1}) \left[ 1-\delta(t_{i}, t_{i+1}) \right] + (N-1)K \nonumber
\end{eqnarray}
where $\beta=1/k_{B}T$ and $\delta(x, y)$ is the Kronecker delta, which equals one if $x=y$ and zero otherwise. Eq.~(\ref{potts}) is a $q$-state Potts-type model, where the generalized spin variables can take values $t=0,1,\dots,q$. For aggregation, the spin states correspond to protein conformations, where $t=0, 1, 2$ indicates that a protein is a random coil, a sheet, or a helical conformation, respectively. The first and second terms are non-zero only when the $ith$ protein is in a sheet or helix conformation, respectively. The free energies described by $P_0$, $P_1$, or $P_2$, refer to the interaction between the $ith$ protein that is coil, sheet, or helical, respectively, and the nearest neighbor protein at location $i+1$. Hence the summation in the first two terms runs to $N-1$ instead of $N$. Even though we think of $P_0$, $P_1$, or $P_2$ as an interaction energy between two neighboring proteins, the energetic weights of these energies are associated with the $ith$ protein, and indeed depend on the conformation of the $ith$ protein. We set the coil interaction energy, $P_{0}$ to zero which serves as a reference for the helix and sheet interactions. Thus, if $P_1 < 0$ ($P_2 < 0$), the random coil interaction is more stable than the sheet (or helix) interaction; if $P_1 > 0$ ($P_2 > 0$), the sheet (or helix) interaction is more stable than the random coil interaction. $K>0$ is an association energy between two monomers that does not depend on conformation. Since $K$ simply links two nearest neighboring monomers, the number of $K$ interactions may be thought of as the degree of polymerization of aggregates. In this way, a dimer composed of two coil monomers will have energy equal to $K$, whereas otherwise dimers would be indistinguishable from monomers.  

\begin{figure}[ht]
	\begin{center}
	\vspace{0.6 cm}
	\includegraphics[width = \columnwidth ]{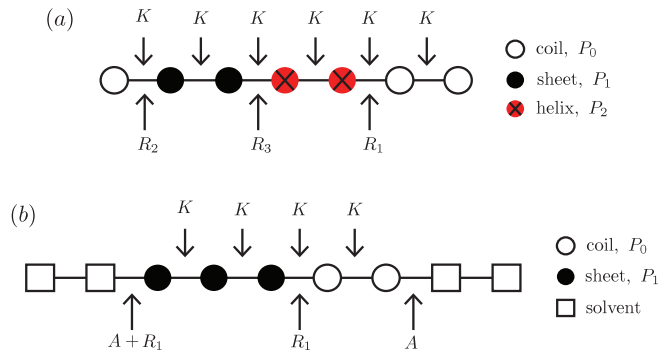}
	\caption{(Color online) Summary of solvent and protein conformation energies. A site is occupied with a solvent, $n=0$ (square), or a protein, $n=1$ (circles). Only proteins may assume a particular conformation (sheet, black/solid circle; helix, red circle marked with X; coil, white circle). In (a), a $q=3$ Potts model for helix-sheet-coil conformations is shown where only protein-protein interactions are illustrated. In (b), a dilute $q=2$ Potts model for sheet-coil conformations is shown where both protein-protein and protein-solvent interactions are indicated, and $n_{c}=1$. In both (a) and (b), interactions that stabilize the aggregate are shown with down arrows, whereas interfacial interactions between different regions of structure are drawn with up arrows.}
	\label{weight}
	\end{center}
	\end{figure}

The third term in Eq.~(\ref{potts}) is a free-energetic penalty associated with the interface between different regions of structure. These interface penalties are parameterized by energies $R_{j}\ge0$, where $j=0$, $1$, or $2$ refers helix-coil, sheet-coil, and helix-sheet boundaries, respectively. The notation $R(t_{i}, t_{i+1})$ refers to the energy of the specific type of boundary: helix-coil or coil-helix boundaries, $R(0,2)=R(2,0) \equiv R_{0}$; sheet-coil or coil-sheet boundaries, $R(0,1)=R(0,1) \equiv R_{1}$; and sheet-helix or helix-sheet boundaries, $R(2,1)=R(1,2) \equiv R_{2}$. Note that the index $j$ in $R_{j}$ does not correspond to $q$ of the Potts model. As to the physical origins of the $R$-terms, they can arise from the effective repulsive interactions between neighboring proteins of different conformations, but more likely, they arise from the loss of entropy at the boundaries between regions of different conformations. $R$ can then be thought of as an initialization parameter, or a barrier to over-come. Overall, six total parameters, which are summarized in Fig.~\ref{weight}(a), are needed to describe possible interactions between proteins. However, in practice it could be less because not all conformations may play significant roles in aggregation. For instance, it is well known that many fibrils are dominated by cross beta-structure, therefore a 2-state, sheet-coil model is a justified system of importance.  

In general, with a simpler two-state system we can model sheet-coil, helix-coil, or even helix-sheet systems using a $q=2$ Potts-type interactions, which can be reduced into an Ising-type model. As an example, let $t_{i}=-1,+1$ refer to whether the $ith$ protein is a random coil or sheet conformation, respectively. The effective Hamiltonian for a Potts model for sheet-coil filaments is 
\small
\begin{equation}
\label{potts_q2}
-\beta \mathcal{H} = P_1 \sum_{i=1}^{N-1} \delta( t_{i}, 1 ) - R_1 \sum_{i=1}^{N-1}  \left[ 1-\delta(t_{i},t_{i+1}) \right] + (N-1)K
\end{equation}
\normalsize
where the coil is taken as the reference state. As with the Potts models~\cite{schreck}, the term $P_1$ corresponds to a ``magnetic-field'' strength, and $R_1$ the spin-spin interaction and the Boltzmann weights $\sigma_1 \equiv \exp(-2R_1)$ and $s_1\equiv \exp(P_1)$ are the Zimm-Bragg-like ``initiation'' and ``propagation'' parameters for sheet-coil protein aggregation~\cite{vander}. By substituting the identity $\delta( t_{i}, t_{j} )$ = $\frac{1}{2}\left( 1 + t_{i}t_{j} \right)$ into Eq.~(\ref{potts_q2}) and simplifying, we get the Ising-type aggregation model of van Gestel et al., Eq.~(2) in Ref.~\onlinecite{dutch_2006}. The only difference in our approach is that we assume a spin variable $t$ refers to a protein conformation whereas in Ref.~\onlinecite{dutch_2006}, $t$ refers to the state of a bond between proteins.

\section{Explicit inclusion of the interactions with solvent}
\subsection{Protein/Solvent interfaces}
Before calculating thermodynamical quantities of aggregates, we consider the effects of solvent on the formation and propagation of protein aggregates. It is generally believed that protein aggregation is a nucleation process~\cite{serio}, where the free energy of a small assembly increases until a nucleus with $n_{c}$ monomers is formed. Creation of nuclei is a slow, stochastic process. Once a nucleus is formed, it may elongate at either end by monomer addition rapidly with free energy going downhill, eventually forming filaments~\cite{morris}. Other more complex pathways are also possible, including the merging of aggregates. Kinetic models are often used to measure the rates of this nucleation/elongation process~\cite{kunes, powers, knowles, kashchiev, collins, morris, muth, cab}. In particular, recent studies by Zhang and Muthukumar~\cite{muth} and others~\cite{cab, schmit} have indicated that nuclei formation occurs only for two, three, $\dots$, $n$-layer aggregates (we call them quasi-1D aggregates), and no nucleation barrier exists in 1D systems. In this section we assume that a nucleus term added to a 1D effective Hamiltonian is a coarse graining of a more realistic quasi-1D model for nuclei, where the lengths of the aggregates are much greater than their widths. This is mainly a simplification, or it can be considered as an approximation to the case where an oligomer is the fundamental unit (or particle) based on which a proto-fibril is formed. On the other hand, as mature amyloid fibrils are known to contain many thousands of proteins and are non-branching structures, they grow primarily in one dimension. More accurately, in Section~\ref{strip}, we will consider the nucleation of a quasi-1D model (or an $n \times N$ model, where $n$=1, 2,$\dots$, and is much smaller than $N$). Comparison with the simpler 1D model in Fig.~\ref{2d_plot} shows that a 1D statistical mechanical model captures some of the features of the $n \times N$ or quasi-1D models.

In our model, the nucleus is a stretch of $n_{c}$ number of sites on a 1D lattice that are occupied by proteins, where aggregates are flanked by solvent on both sides, and the proteins are linked via the interaction $K$, discussed above. The solvent could be, for example, a cluster of water molecules. We assume that the nucleus surrounded by solvent define an interface that is described by the free-energy $A\ge0$. This interface energy may be attributed to surface tension between solvent and a nucleus, where proteins in the nucleus may be in contact and are involved in long-range interactions.  At each site on the lattice, the occupation variables $n_{i}=0,1$ indicate whether the site is occupied by solvent or protein, respectively. The 1D lattice, on the other hand, can be considered to be embedded in a large overall space, either 2D or 3D, which is filled with solvent and a dilute protein solution. That is, we assume the distance between any two aggregates is large enough that we can focus on a single one at a time. In some way our approach is similar to other spin-models for aggregation that put solvent and protein on the lattice~\cite{wu, spin_one}.

To include the nucleus-solvent interfacial free-energy, we modify Eq.~(\ref{potts_q2}). To ease notation, define $\chi( x, y ) \equiv 1 - \delta( x, y )$ where $x$, $y$ can be either spin or occupation variables located at sites $i$ and $j$. $\chi$ is zero if $x=y$ and 1 otherwise. The lattice-gas effective Hamiltonian for interactions between sheet or coil proteins as well as nuclei-solvent interfaces on a 1D lattice with $N_{T}$ sites is now given by
\small
\begin{eqnarray}
\label{latticegas}
-\beta \mathcal{H}_{fil} &=& -\beta \mathcal{H}_{pp} - \beta \mathcal{H}_{ps}^{n_c} \\
\label{pro_pro}
-\beta \mathcal{H}_{pp} &=& \sum_{i=1}^{N_{T}-1} \left\{ P_1~\delta( t_{i}, 1 ) + K - R_1\chi(t_{i},t_{i+1}) \right\} n_{i} n_{i+1} \nonumber \\
&-& \sum_{i=1}^{N_{T}-1} R_1\chi(n_{i}, n_{i+1}) \left[ \delta(t_{i},1) n_{i} + \delta( t_{i+1}, 1) n_{i+1} \right] \\
-\beta \mathcal{H}_{ps}^{n_{c}} &=& - \sum_{i=1}^{N_{T}-n_{c}-1} A \chi( n_i, n_{i+n_{c}} ) \prod_{j=i+1}^{i+n_{c}-1} \delta( n_j, 1 ) 
\label{nucleus}
\end{eqnarray}
\normalsize
where `pp' in $-\beta \mathcal{H}_{pp}$ refers to `protein-protein' interactions and `ps' in $-\beta \mathcal{H}_{ps}^{n_{c}}$ refers to `protein-solvent' interactions. Eq.~(\ref{nucleus}) is the effective Hamiltonian associated with a nuclei-solvent interface, with $\chi( n_i, n_{i+n_{c}} )$ ensuring that there is solvent at site $i$ and a protein at $i+n_{c}$, or vice-versa. The product of Kronecker terms fixes all the remaining sites between the solvent at $i$ and the protein at $i+n_{c}$ to be occupied by proteins. In Eq.~(\ref{latticegas}), terms with $P_{1},~K,$ and $R_{1}$ have the same meaning as in Eq.~(\ref{potts_q2}) and make up the effective Hamiltonian for sheet-coil filaments in the lattice-gas Potts model. Now $K$ explicitly depends on whether two neighboring sites are occupied by proteins and facilitates the elongation of an aggregate.  $A$ is the nucleus-solvent interfacial free-energy. To avoid introducing any more free parameters, we assume in second summation in Eq.~(\ref{pro_pro}) that the interaction between a sheet protein immediately flanked by solvent is described by the interaction free-energy $R_1$. With this convention, both ends of a sheet segment contribute a factor of $R_1$, regardless of whether the segment is flanked by proteins or solvent. The free energies for sheet-coil aggregates including solvent is summarized in Fig.~\ref{weight}(b). 

Eq.~(\ref{latticegas}) is a more general approach to fibril elongation when compared to previous statistical mechanical models for protein aggregation~\cite{nyrkova, dutch_2001, dutch_2006, lee, schmit}, which focus on specific aggregation pathways. Fibrils may grow longer via monomer addition at fibril ends, which in a sense is similar to some kinetic models for elongation, in particular, the model proposed by Massi and Straub~\cite{massi}. Additionally, by using the lattice-gas formalism, Eq.~(\ref{latticegas}) can accommodate a variety of elongation mechanisms including merging and fracturing of aggregates of different sizes along the 1D lattice. In reality the merging of filaments and proto-fibrils is not a 1D process, and in Section~\ref{strip} we will consider a related effective Hamiltonian on a strip lattice to model interactions between 1D filaments. In this section, we focus on a 1D model for aggregate elongation. 

\subsection{Average properties and thermodynamics}
\label{ave_thermo}
Now that we have discussed interactions between protein-protein and protein-solvent, we can calculate thermodynamic quantities and test model predictions against experimental data. First, we must calculate the partition function for Eq.~(\ref{latticegas}). Since the number of proteins on the 1D lattice may fluctuate, we work within the grand canonical ensemble where $N_{T}$ refers to the total number of lattice sites, and $N_p$=$\sum_{i=1}^{N_{T}}n_{i}$ refers to the total number lattice sites occupied by proteins. $\mathcal{Q}$ is a grand partition function. Substituting Eqs.~(\ref{pro_pro}) and ~(\ref{nucleus}) into Eq.~(\ref{latticegas}), we write $\mathcal{Q}$ for the lattice-gas filament model as 
\begin{equation}
\label{grand}
\mathcal{Q} = \sum_{\{t\}, \{n\}} \exp\left(-\beta \mathcal{H}_{fil} + \beta \mu_{PC} N_{p}\right) 
\end{equation}
where \textcolor{black}{$\beta \mu_{PC}$ is the dimensionless chemical potential arising from the contact and interface interactions between proteins in aggregates}, and the notation $\{t\}, \{n\}$ means summation over the spin, occupancy variables, respectively, at each site. For $N_{T}>2n_{c}$, $\mathcal{Q}$ may be solved for exactly by a transfer matrix $T$ as, 
\small
\begin{eqnarray}
\label{pffactor}
\mathcal{Q} &=& \sum_{\{t\}, \{n\}} \prod_{i=1}^{N_{T}-n_{c}} \mathcal{T} (t_{i}, t_{i+1}, n_{i}, n_{i+1}, \dots, n_{i+n_{c}}) \\ 
\label{transfer}
\mathcal{T} &=& \exp \left\{ \left[ P_1 \delta( t_{i}, 1 ) + R_1 \chi( t_{i}, t_{i+1} )  + K \right] n_{i} n_{i+1}   \right\} \nonumber \\ 
&\times& \exp \left\{ A \chi( n_i, n_{i+n_{c}} ) \prod_{j=i+1}^{i+n_{c}-1} \delta( n_j, 1 ) \right\}\\ 
&\times& \exp \left\{ R_1\chi(n_{i}, n_{i+1}) \left[ \delta(t_{i},1) n_{i} + \delta( t_{i+1}, 1) n_{i+1} \right]  + \beta \mu_{PC} n_{i}\right\}  \nonumber 
\end{eqnarray}
\normalsize
where we sum over conformations only if the $ith$ site is occupied by a protein, i.e., $n_{i}=1$. Notice that the parameter for sheet propagation is counted only when the $ith$ site is a sheet protein. As an explicit example in writing out the transfer matrix, we consider the case $n_{c}=1$ for a two-state system. Using Eq.~(\ref{transfer}) gives the elements of the following matrix
\begin{eqnarray}
\mathcal{T} &=&
\begin{tabular}{ c c  | c c c}
	& $t_{i+1}$ &  & $-1$ & $1$ \\
	& $n_{i+1}$ & $0$ & $1$ & $1$ \\
	$t_{i}$ & $n_{i}\diagdown$ &  &  &  \\
  	\hline\noalign{\smallskip}
   	 & $0$~~ & $1$ & $\sqrt{\alpha}$ & $\sqrt{\alpha \sigma_1}$ \\
  	$-1$ & $1$~~ & $z\sqrt{\alpha}$ & $kz$  & $kz\sqrt{\sigma_1}$ \\ 
  	~$1$ & $1$~~ & $z\sqrt{\alpha \sigma_1}$ & $kzs_{1}\sqrt{\sigma_1}$ & $kzs_{1}$  \\
\end{tabular}
\end{eqnarray}
where $s_1$ and $\sigma_1$ were defined above, $k\equiv\exp(K)$ and $\alpha\equiv\exp(-2A)$ are the Zimm-Bragg-like parameters, and $z \equiv \exp(\beta \mu_{PC})$. The matrix elements $\mathcal{T}_{i,j}$ represent the probability of each type of interaction. For general $q$ and $n_{c}$, the transfer matrix has dimension $(q+1)^{n_{c}} \times (q+1)^{n_{c}}$ and $N_{\lambda}=(q+1)^{n_{c}}$ number of eigenvalues. 

Now we can write $\mathcal{Q}$ and calculate thermodynamic properties using the eigenvalues of the transfer matrix. For a finite lattice boundary conditions must be specified. These could be open, where either ends of the lattice could be occupied by a protein of a specified conformation or solvent, or periodic, where the lattice simply forms a ring. For any case, we have
\begin{equation}
\mathcal{Q} = \sum_{i=1}^{N_{\lambda}} x_{i} \lambda_{i}^{N_{T}-n_{c}}
\end{equation}
where the coefficients $x_{i}$ are determined by the specified boundary conditions. If periodic boundary conditions are imposed, we set $t_{N_{T}-n_{c}+1}=t_{1}$, $t_{N_{T}-n_{c}+2}=t_{2}$, $\dots$, $t_{N_{T}}=t_{n_{c}}$ so that all coefficients $x_{i}$ are unity and the partition function is found easily from 
\begin{eqnarray}
\mathcal{Q}&=&Tr(T^{N_{T}}) = \lambda_1^{N_T} \left( 1+ \sum_{i=2}^{N_{\lambda}} \left(\frac{\lambda_i}{\lambda_1} \right)^{N_T}  \right) \\ 
&\approx& \lambda_1^{N_T}
\label{large}
\end{eqnarray}
where $Tr$ is the trace operation, $\lambda_1$ is the largest eigenvalue of the transfer matrix, $\lambda_2$ is the second largest eigenvalue of the transfer matrix, and so on. Eq.~(\ref{large}) is valid when the lattice grows large and in the thermodynamic limit $N_{T} \to\infty$, 
\begin{equation}
\label{thermo_soln}
N_{T}^{-1} \ln \mathcal{Q} = \ln \lambda_{1}.
\end{equation}

Finally, we calculate some properties of the system. \textcolor{black}{Of particular interest are the average number of proteins on the lattice, $\langle N_p \rangle$, which we refer to as the occupation of the lattice}, the number of proteins in filaments, $\langle \psi \rangle $, the number of filaments, $\langle \gamma \rangle $, the number of sheet proteins in filaments, $\langle \theta \rangle$, and the number of sheet segments, $\langle \nu \rangle $, as 
\begin{eqnarray}
\label{therm1}
\langle N_{p} \rangle &\equiv& z \frac{\partial}{\partial z} \ln \mathcal{Q} \\ 
\label{therm3}
\langle \gamma \rangle  &\equiv& \frac{1}{2} \frac{\partial}{\partial A} \ln \mathcal{Q} \\ 
\label{therm2}
\langle \psi \rangle  &\equiv& \frac{\partial}{\partial K} \ln \mathcal{Q} + \langle \gamma \rangle  \\ 
\label{therm5}
\langle \nu \rangle  &\equiv& \frac{1}{2} \frac{\partial}{\partial R_1} \ln \mathcal{Q} \\ 
\label{therm4}
\langle \theta \rangle &\equiv&  \frac{\partial}{\partial P_1} \ln \mathcal{Q} 
\end{eqnarray}
respectively. In each expression all energies except the varying one are held constant upon differentiation. A factor of $1/2$ in Eqs.~(\ref{therm3}) and (\ref{therm5}) corrects the over-counting of the number of distinct filaments and extended sheet regions. We also calculate the average length of aggregates, $\langle L_{p} \rangle$, and the average length of sheet segments, $\langle L_{s} \rangle$, according to
\begin{eqnarray}
\label{polylength}
\langle L_{p} \rangle &\equiv&  \frac{\langle \psi \rangle }{\langle \gamma \rangle  + 1}  \\ 
\label{sheetlength}
\langle L_{s} \rangle &\equiv&  \frac{\langle \theta \rangle}{\langle \nu \rangle  + 1}  
\end{eqnarray}
respectively. The factor of 1 in Eqs.~(\ref{polylength}) and (\ref{sheetlength}) accounts for the case where proteins completely occupy the lattice.

\subsection{Numerical Results}
	\begin{figure}
	\vspace{0.6 cm}
	\begin{center}
	\includegraphics[width = \columnwidth]{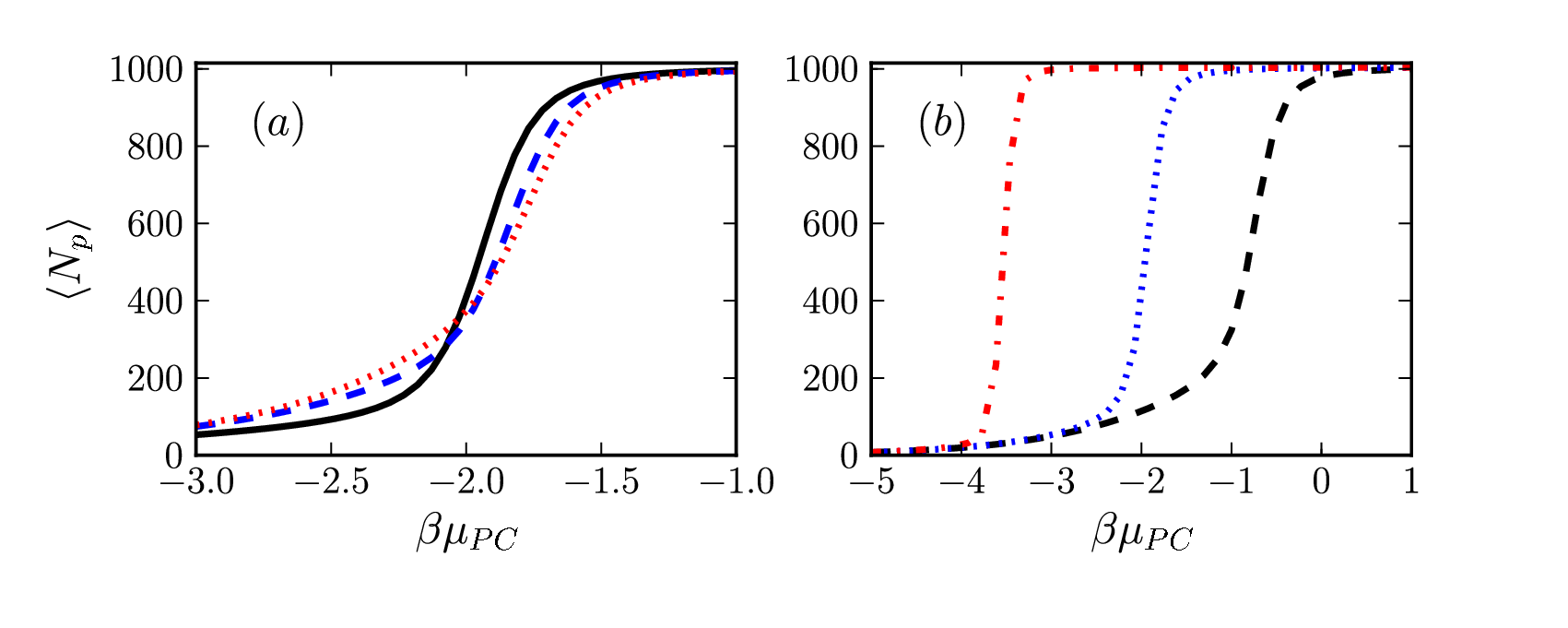}  
	\caption{(Color online) Plot (a) illustrates the effect of varying the contact chemical potential, $\beta \mu_{PC}$, on protein number, $\langle N_p \rangle$, with $n_{c}$=2 (solid, black), 4 (dashed, blue), 6 (dotted, red). In (b), $\langle N_p \rangle $ vs. $\beta \mu_{PC}$ is shown for $K$= 0$k_{B}T$ (solid, black), $1k_{B}T$ (dashed, blue) and $2.5k_{B}T$ (dotted, red). Unless otherwise stated, $K=P=1k_{B}T$, $R=A=1k_{B}T$, $N_{T}=1000$, and $n_{c}=2$.  }
	\label{cover}
	\end{center}
	\end{figure}

	\begin{figure}
	\vspace{0.6 cm}
	\begin{center}
	\includegraphics[width = \columnwidth]{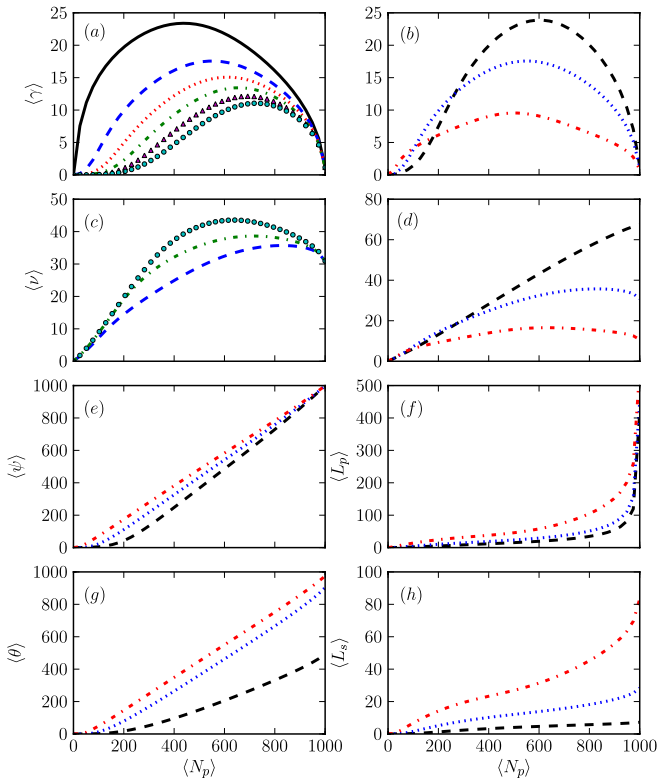}  
	\caption{(Color online) Plot (a) illustrates the effect of varying $n_{c}$ and protein number, $\langle N_p \rangle$, on $\langle \gamma \rangle $, i.e., Eq.~(\ref{therm3}), with $n_{c}$=1 (solid, black), 2 (dashed, blue), 3 (dotted, red), 4 (dashed-dotted, green), 5 (purple triangles), 6 (cyan circles). In (b) $\langle \gamma \rangle $ vs. concentration is show for $K$= 0$k_{B}T$ (dashed, black), $1k_{B}T$ (dotted, blue) and $2.5k_{B}T$ (dotted-dashed, red). Plot (c) illustrates the effect of varying $n_{c}$ and $\langle N_p \rangle$ on $\langle \nu \rangle $, i.e., Eq.~(\ref{therm5}), with $n_{c}$=2 (dashed, blue), 4 (dashed-dotted, green), 6 (cyan circles). In (d) $\langle \nu \rangle $ vs. $\langle N_p \rangle$ is shown for $P$= $0k_{B}T$ (dashed, black), $1k_{B}T$ (dotted, blue) and $2k_{B}T$ (dotted-dashed, red). In (e) and (f), $\langle \psi \rangle $ and $\langle L_{p} \rangle$ are plotted against $\langle N_p \rangle$, respectively, where in both plots $K$= $0k_{B}T$ (dashed, black), $1k_{B}T$ (dotted, blue) and $2.5k_{B}T$ (dotted-dashed, red). Finally, In (g) and (h), $\langle \theta \rangle$ and $\langle L_{s} \rangle$ are plotted against $\langle N_p \rangle$, respectively, where in both plots $P_1$= $0k_{B}T$ (dashed, black), $1k_{B}T$ (dotted, blue) and $2k_{B}T$ (dotted-dashed, red). Unless otherwise stated, $K=P=1k_{B}T$, $R=A=1k_{B}T$, $N_{T}=1000$, and $n_{c}=2$.  }
	\label{allplot}
	\end{center}
	\end{figure}


\textcolor{black}{In this section we compute the thermodynamic quantities represented by Eqs.~(\ref{therm1}-\ref{sheetlength}) for varying system parameters and different $n_c$. For later use, we define the normalized number of proteins on the lattice (referred to as the coverage) as 
\begin{equation}
\label{con}
\phi\equiv\frac{\langle N_p \rangle}{N_{T}}.
\end{equation}
$\phi$ can also be thought of as the concentration of proteins in the aggregate phase. }

\textcolor{black}{In Fig.~\ref{cover}, we plot the average number of proteins on the 1D lattice, $\langle N_p \rangle$, versus the chemical potential contribution from the contacts, $\mu_{PC}$. The values of $\mu_{PV}$, $\mu_{ST}$, and $\mu_{SR}$ in Eqs.~(\ref{chemsoln}) and (\ref{chemagg}) are regarded as constants at a specified temperature. Thus varying $\mu_{PC}$, through Eq.~(\ref{equib}), can be accomplished by changing the experimental concentration, $c$, of protein in solution. Both Fig.~(\ref{cover}) (a) and (b) illustrate the dependence of $\langle N_p \rangle$ versus $\mu_{PC}$, where for large, negative values of $\mu_{PC}$, almost no proteins are found on the lattice and in aggregates. In other words, at low protein solution, $c$, aggregates are found in extremely few numbers. As protein concentration, $c$, increases (i.e. $\mu_{PC}$ increases), proteins may form aggregates in greater numbers, and at an increasing rate as the lattice becomes nearly half saturated. Further increasing the protein concentration in solution allows more monomers to join aggregates rather easily until the lattice becomes saturated. In Fig.~\ref{cover} (a), the effect of varying the critical concentration on average number of proteins, $\langle N_p \rangle$, is illustrated, where increasing $n_{c}$ is seen to have only a marginal effect on the $\langle N_p \rangle$ dependence on $\mu_{PC}$. Whereas in Fig.~\ref{cover} (b), varying system parameters that parametrize the contact strengths clearly influences the average number of proteins on the lattice at particular experimental concentrations. For instance, as illustrated in Fig.~\ref{cover} (b), increasing the strength of interactions between proteins, $K$, causes proteins to join aggregates at lower concentrations of monomers in solution.}

In Figs.~\ref{allplot}(a-h) we considered a 2-state sheet-coil model on a finite lattice with $N_{T}=1000$ total sites and periodic boundary conditions imposed. In Fig.~\ref{allplot}(a), we show effects of varying $n_{c}$. As protein occupation, $\langle N_p \rangle$, increases, the number of filaments increases to a maximum value, then, the filament numbers decrease with $\langle N_p \rangle$ as the lattice becomes saturated with proteins. Increasing $n_{c}$ from 1 to 6 progressively increases the value of $\langle N_p \rangle$ for both the onset of filament nucleation and the maximum number of filaments, respectively, and also decreases filament numbers overall for all values of $\langle N_p \rangle$. As shown in Fig.~\ref{allplot}(b), increasing the association energy between monomers, $K$, from $0k_{B}T$ shifts the value of $\langle N_p \rangle$ where $\langle \gamma \rangle $, the number of filaments, reaches a maximum to lower values. Additionally, increasing $K$ causes $\langle \gamma \rangle$ to rise faster at low protein average number of proteins, while also progressively reducing the overall number of filaments at values of $\langle N_p \rangle$ away from zero. 

In Fig.~\ref{allplot}(c) and (d), we plot the number of sheet segments, $\langle \nu \rangle $, versus the protein occupation, $\langle N_p \rangle$, for various $n_{c}$ (Fig.~\ref{allplot}(c)) and $P_1$ (Fig.~\ref{allplot}(d)). The number of sheet segments, $\langle \nu \rangle $, increases with $\langle N_p \rangle$ until reaching a maximum, then decreases toward a common value at maximum protein occupation $\langle N_p \rangle$=$N_T$. In Fig.~\ref{allplot}(c), increasing $n_{c}$ increases the maximum number sheet segments since larger nuclei may contain more sheet-coil interfaces than smaller nuclei. Also, the maximum of $\langle \nu \rangle $ occurs at progressively lower protein occupation as $n_{c}$ increases. Fig.~\ref{allplot}(d) shows that increasing the interaction strength between sheet proteins, $P_1$, reduces the total number of sheet segments for all but the lowest values of protein occupation, while increasing the average length of the sheet segments (see Fig.~\ref{allplot}(h)). The maximum value for $\langle \nu \rangle $ is also achieved at lower protein occupations for increasing $P_1$. 

In Fig.~\ref{allplot}(e), we plot the number of proteins in filaments, $\langle \psi \rangle $, versus protein occupation, $\langle N_p \rangle$. In Fig.~\ref{allplot}(f), we plot the average length of aggregates, $\langle L_{p} \rangle$, versus protein occupation. In both figures $K$ is varied as well. As protein occupation of the lattice increases, proteins start to join filaments, and $\langle \psi \rangle$ increases almost linearly with $\langle N_p \rangle$. The lengths of filaments also increase as proteins join filaments, but not linearly. Once the lattice becomes occupied mostly by proteins, the lengths take off and reach a maximum value at high protein occupation. Thus, increasing $K$ increases the numbers of proteins in filaments and the lengths of the filaments. 

Finally, we plot the number of sheet proteins in filaments, $\langle \theta \rangle$, and the length of sheet segments, $\langle L_{s} \rangle$, versus protein occupation, $\langle N_p \rangle$, in Fig.~\ref{allplot}(g) and (h), respectively, for different $P_1$ values. The behaviors of  $\langle \theta \rangle$ and $\langle L_{s} \rangle$ are similar to the behaviors of $\langle \psi \rangle $ and $\langle L_{p} \rangle$, while increasing $P_1$ clearly increases the number of sheet proteins and the sheet segment lengths. Varying $n_{c}$ only marginally changes $\langle \psi \rangle $, $\langle L_p \rangle$, $\langle \theta \rangle$, and $\langle L_s \rangle$ (not shown). 

	\begin{figure}
	\begin{center}
	\includegraphics[width = \columnwidth ]{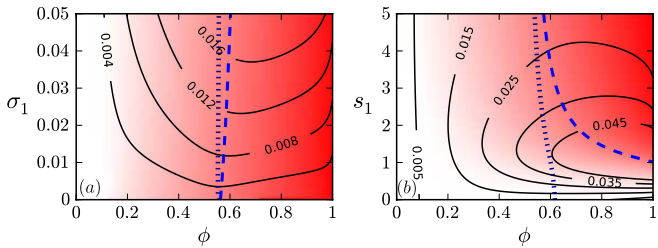}
	\caption{(Color online) The phase plots of the number of sheet segments, $\langle \nu \rangle$ (solid, black lines), and the number of proteins in sheet segments, $\langle \theta \rangle$, versus system parameters (a)  $\sigma_1=\exp(-2R_1)$ and (b) $s_1=\exp(P_1)$, all versus coverage, $\phi$. In both plots, normalized $\langle \theta \rangle$ at a particular ($\sigma_1$, $\phi$) may vary from zero to one, with white color indicating $\langle \theta \rangle=0$ and solid red indicating $\langle \theta \rangle=1$. The white-to-red gradation represents values in the range 0-to-1. Dashed (blue) lines indicate where filaments contain equal parts sheet and coil, which we define to be the locations of gradual, conformational phase transitions. Regions to the left of dashed lines indicate filaments are mostly composed of coils, whereas regions to the right of the dashed line indicate the filaments have majority sheet structure. Also, regions to the left of the dotted (dark blue) line indicate more solvent than proteins on the lattice, whereas regions to the right of the dotted line indicate more proteins than solvent on the lattice. We refer to the dotted line as the locations of solvent/protein equal-population. Unless otherwise indicated in the plots, $P_{1}=1k_{B}T$, $K=1k_{B}T$, $R_{1}=1k_{B}T$, $A=1k_{B}T$ and $n_{c}=2$.}
	\label{sigphase}
	\end{center}
	\end{figure}

In addition to quantities plotted in Fig.~\ref{allplot}, we present phase diagrams in which thermodynamic properties of aggregates are plotted as functions of interaction parameters. These plots yield information on when sheet and coil proteins are in equal numbers, locations we define as sheet-coil phase transitions of filaments. In Fig.~\ref{sigphase}(a), the number of sheet segments, $\langle \nu \rangle $, and the number of sheet proteins in aggregates, $\langle \theta \rangle$, vs. $\sigma_{1}$ and $\phi$ are computed, respectively, and in Fig.~\ref{sigphase}(b), $\langle \nu \rangle $ and $\langle \theta \rangle$ vs. $s_{1}$ and $\phi$ are shown.

In Fig.~\ref{sigphase}(a) the maximum number of sheet segments occurs at high protein coverage and weak sheet-coil interface interactions, that is $\sigma_{1}\approx0.05$. From this region of the phase plot, $\langle \nu \rangle $ decreases in every direction, which means sheet segments decrease in numbers for smaller protein coverage, and also when the interaction energy of a sheet-coil interface increases, i.e., $\sigma_{1}<<0.05$. The number of sheet proteins in filaments, $\langle \theta \rangle$ is maximal at high protein coverage, and decreases in magnitude eventually tending toward $\langle \theta \rangle\approx0$ as the protein coverage decreases. However, at high protein coverage, the lengths of sheet segments (not shown) are longest when the sheet-coil interface interaction is large, $\sigma_{1}\approx0$, and shortest when the interaction is small, $\sigma_{1}\approx0.05$. Additionally, the curve representing equal numbers of solvent and protein on the lattice (referred to as the `solvent/protein' curve) is not strongly dependent on the value of $\sigma_{1}$. However, coil-sheet transition locations tend toward higher protein coverage as the sheet-coil interface energy weakens and eventually $\sigma_1 \approx 0.05$.

In Fig.~\ref{sigphase}(b), the number of sheet segments, $\langle \nu \rangle $, is maximal at high protein coverage and also when $s_{1}\approx1$ where the interactions between sheet proteins are weak or zero. $\langle \theta \rangle$ is maximal at high protein coverage and large interactions between sheets, i.e., large $s_1$, and decreases in every direction from this region. The solvent/protein curve location occurs at essentially a fixed protein coverage for $s_{1}>1$, but for $s_{1}<1$, the curve tends slightly toward higher protein coverage. On the other hand, for large $s_{1}$, the coil-sheet transition occurs at roughly the same protein coverage (about $\phi = 1/2$), but once $s_{1}$ decreases towards $s_{1}=1$, the protein coverage where coil-sheet transition occur increases, tending toward $s_{1}\approx1$ at very high protein coverage. Thus, once $s_{1}<1$, interactions between sheet proteins are repulsive and the proteins in filaments are largely in coil conformations. However, this region may be unphysical as large aggregates of proteins are known to contain $\beta$-structure. 

\section{Quasi-1D models for protofibrils and fibrils}
\label{strip}
\begin{figure}
	\begin{center}
	\includegraphics[width = \columnwidth]{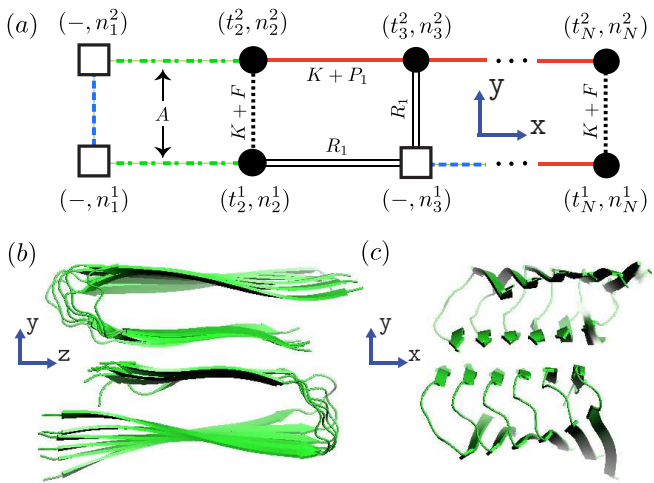} 
	\caption{(Color online) (a) Graphical illustration of Eq.~(\ref{hamilb}) on a $2 \times N$ strip lattice. A black dot indicates that a vertex is occupied by a sheet protein, a white square indicates solvent. Solid red lines indicate interactions between proteins along the x-axis, while dotted black lines are interactions between two sheet proteins on the y-axis. Dashed-dotted green lines indicate a boundary for a nucleus, which is the dimer ($n_{c}=2$) positioned on the y-axis. Dashed blue lines indicate no interaction between connected vertices. Double solid lines are sheet-solvent interfaces. (b) Front-view (y-z plane) of an aggregate of A$\beta$(1-40) proteins. (c) Side-view (x-y plane) of A$\beta$(1-40) proteins illustrating the inter-filament interactions~\cite{petkova}.}
	\label{ladder}
	\end{center}
	\end{figure}

Protein protofibrils and fibrils comprise of several filaments. To study thermodynamic properties of fibrils or proto-fibrils, we add the interaction energy terms between $L_y$ number of filaments in the effective Hamiltonian and put the fibrils onto a $L_y \times N$ strip lattice, which is a finite strip in one direction of an $N \times N $ square lattice. The $2 \times N$ strip is illustrated in Fig.~\ref{ladder}(a). In Fig.~\ref{ladder}(b) and (c) the representation of a proto-filament of A$\beta$(1-40) is shown, originally produced Tycho and coworkers~\cite{petkova}. We will model this proto-filament as two 1D filament-like structures that propagate in the x-direction, as indicated in Fig.~\ref{ladder}(b) and (c). In our model the proto-filament could grow either by joining two filaments together, or as a quasi-1D aggregate growing from a single nuclei arranged on the y-axis. \textcolor{black}{Of course, our statistical mechanical models deal only with equilibrium properties, not kinetic mechanisms of fibril growth.}

The position of a protein or solvent is represented by a vertex within the strip, and is specified by coordinates ($i,~j$), which are the positions on the x and y-axis, respectively, of the strip. The total number of vertices is $N_{TOT}=L_{y} N$. The strip lattice in Fig.~\ref{ladder}(a) contains spin and lattice-gas variables $t_{i}^{j}$ and $n_{i}^{j}$, respectively, at each vertex ($i$, $j$). The spin variables $t_{i}^{j}$ = $0, 1, \dots, q$, represent different conformation states of a protein and $n_i^j$=$0,1$ denotes lattice gas or occupation states. For simplicity we assume that interactions between neighboring proteins on the y-axis are restricted to vertices that have the same index $i$, and the proteins occupying these sites must both be locked in the sheet conformation. We consider the strip that is composed of two identical 1D lattices aligned in register, but this does not mean the filaments have to be in register since we allow the number of proteins to fluctuate. This is a main difference of our model from the simpler method of counting inter-filaments and loose ends in the model of van Gestel~\cite{dutch_2006, dutch_2008}. 

The inter-filament interactions between two 1D filaments are treated using a model similar to the 2-helix chain model proposed by Skolnick~\cite{skolnick} and others~\cite{qian, ghosh_dill} which uses ZB parameters for describing the inter-residue interactions between two independent $\alpha$-helical protein chains. In general, the Hamiltonian for an $L_y \times N$ strip lattice that includes inter-filament interactions is written using the 1D Hamiltonian, Eq.~(\ref{latticegas}), by changing the spin and lattice-gas variables $t_{i} \rightarrow t_{i}^{j}$ and $n_{i} \rightarrow n_{i}^{j}$, respectively, as
\begin{eqnarray}
\label{two_filament}
-\beta \mathcal{H}_{strip}^{A} &=& - \sum_{j=1}^{L_y} \beta \mathcal{H}_{fil}(j)  \\
&+& F \sum_{i=1}^{N} \sum_{j=1}^{L_y-1} \delta(t_{i}^{j}, 1) \delta(t_{i}^{j+1}, 1) n_{i}^{j}n_{i}^{j+1} \nonumber
\end{eqnarray}
where the notation $\mathcal{H}_{fil}(j)$ refers to the $jth$ filament. For A$\beta$(1-40), we take $L_{y}=2$, illustrated in Fig.~\ref{ladder}(b) and (c). $F$ parametrizes the interaction energy between two sheet-linked proteins which have the same $ith$ index. That is to say, residues from neighboring filaments that are close in real space participate in stabilizing interactions between filaments. In our treatment $F>0$, the proto-fibrils and fibrils are more stable than single filaments.

On the other hand it is known that nucleation does not occur in a truly 1D system~\cite{muth}, so we consider a similar model for aggregates that positions the nucleus along the y-axis as shown in Fig.~\ref{ladder}(a). From this point of view the orientations of proteins in the nucleus are perpendicular to the direction of propagation (x-axis) of the fibrils, and the nucleus is now a multi-layer, quasi-1D structure on a $L_y \times N_T$ ladder. This characterization of the nucleus corresponds with the findings of Zhang and Muthukumar~\cite{muth} that the nucleus contains at least two layers of $\beta$-sheet. The nuclei will assemble into proto-fibrils that grow longer on the quasi-1D lattice. An effective Hamiltonian for quasi-1D aggregation including the multi-layer nucleus term can be written 
\small
\begin{eqnarray}
\label{hamilb}
-\beta \mathcal{H}_{strip}^{B} &=& - \sum_{j=1}^{L_y} \beta \mathcal{H}_{pp}(j) - \sum_{j=1}^{L_y-1} \beta H_y(j) \\ 
- \beta H_y(j) &=& \sum_{i=1}^{N_{T}} \left\{ F~\delta( t_{i}^{j}, 1 )+ K - R_1\chi(t_{i}^{j} ,t_{i}^{j+1})\right\} n_{i}^{j} n_{i}^{j+1} \nonumber \\
&-& \sum_{i=1}^{N_{T}} R_1\chi(n_{i}^{j} , n_{i}^{j+1}) \left[  \delta(t_{i}^{j} ,1) n_{i}^{j} + \delta( t_{i}^{j+1}, 1) n_{i}^{j+1} \right]  \nonumber \\ 
&-& \sum_{i=1}^{N_T - 1} A \prod_{j=1}^{L_y - 1} \chi(n_{i}^{j} , n_{i+1}^{j})
\end{eqnarray}
\normalsize
where the term $-\beta H_{pp}(j)$ given by Eq.~(4) is, upon changing the spin and lattice-gas variables $t_{i} \rightarrow t_{i}^{j}$ and $n_{i} \rightarrow n_{i}^{j}$, respectively, the $jth$ effective Hamiltonian for a 1D filament in the x-direction, one for each layer of the strip lattice. In the y-direction we write analogous interactions, $-\beta H_y$, similar to that in the x-direction, except we introduce $F$ to represent interactions between two sheet proteins. Also included in the y-direction is the nucleus term containing the parameter $A$, which has the same meaning of surface energy as before. 

For both cases the total number of proteins on a strip lattice is then $N_{strip}\equiv\sum_{i}^{N} \sum_{j}^{L_y} n_{i}^{j}$ so that the grand partition function is 
\begin{equation}
\label{strip_pf}
\mathcal{Q}_{strip}^{A(B)} = \sum_{\{t\}, \{n\}} \exp\left(-\beta \mathcal{H}_{strip}^{A(B)} + \beta \mu_{PC} N_{strip}\right) 
\end{equation}
where the sums over ${\{t\}, \{n\}}$ are for all $i$ and $j$, and A, B refers to the effective Hamiltonians given by Eqs.~(\ref{two_filament}) or (\ref{hamilb}), respectively. The grand partition function is solved as $\mathcal{Q}_{strip}^{A(B)}=Tr\left(T_{strip}^{A(B)} \right)^N$ where $T_{strip}^{A(B)}$ is now the transfer matrix that relates nearest-neighbor spin variables $t_{i}^{j}$, $t_{i+1}^{j}$, $t_{i}^{j+1}$, $t_{i+1}^{j+1}$ and lattice-gas variables $n_{i}^{j}$, $n_{i+1}^{j}$, $n_{i}^{j+1}$, $n_{i+1}^{j+1}$. Just as in Section~\ref{ave_thermo}, in the thermodynamic limit $N_T \to\infty$,
\begin{equation}
\label{strippf}
\left(L_y N_{T}\right)^{-1} \ln \mathcal{Q}_{\text{strip}}^{A(B)} = \ln \lambda_{1}^{A(B)}
\end{equation} 
where $\lambda_{1}^{A(B)}$ is the largest eigenvalue of $T_{strip}^{A(B)}$. In general, the dimension of the transfer matrix $T_{strip}^{A}$ is $(q+1)^{n_c L_y} \times (q+1)^{n_c L_y}$ and has $(q+1)^{n_c L_y}$ number of eigenvalues, whereas the transfer matrix $T_{strip}^{B}$ is $(q+1)^{L_y} \times (q+1)^{L_y}$ and has $(q+1)^{L_y}$ number of eigenvalues.

	\begin{figure}
	\begin{center}
	\includegraphics[width = \columnwidth]{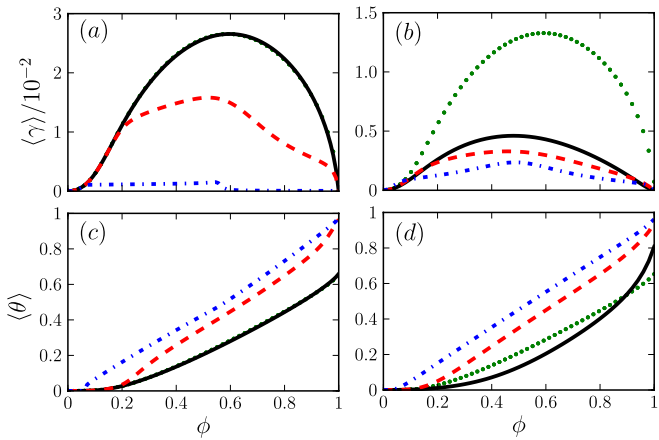} 
	\caption{(Color online) The protein coverage $\phi$ is plotted against the number aggregates, $\langle \gamma \rangle$, in (a) for model A and (b) for model B. The total number of sheet proteins in aggregates, $\langle \theta \rangle$, is plotted in (c) for model A and (d) for model B. Green circles in (a) and (c) are the results of the 1D model for $\langle \gamma \rangle$, whereas in (b) and (d) green circles denote the results of the 1D model for $\langle \theta \rangle$. In all cases, $P_1$ = $0.25k_{B}T$, $K$ = $1k_{B}T$, $A$ = $1k_{B}T$, $R_1$ = $1k_{B}T$.  In all plots, the case $F$ = $0k_{B}T$ are solid black lines, $F$ = $1k_{B}T$ are dashed red lines, and $F$ = $3k_{B}T$ are dashed-dotted blue lines.}
	\label{2d_plot}
	\end{center}
	\end{figure}
The normalized average number of sheet proteins for either case A or B is calculated by substituting Eq.~(\ref{strip_pf}) into Eq.~(\ref{therm4}) and dividing by $L_y N_T$. Additionally, the normalized number of sheet interactions in the y-direction is given by
\begin{eqnarray}
\label{thetay}
\langle \theta_y \rangle &\equiv&  \frac{L_y}{L_y-1}\frac{\partial}{\partial F} \ln \lambda_{1}^{A(B)} 
\end{eqnarray}
for either case A or B. Additionally, the number of aggregates on the strip, $\langle \gamma \rangle$, is found by substituting Eq.~(\ref{strippf}) into Eq.~(\ref{therm3}) and normalizing with respect to $L_y N_{T}$. $\langle \psi \rangle$ now yields the total polymerization of aggregates on the strip lattice, but it does not yield the correct number of proteins in aggregates. Additionally, $\langle \nu \rangle$ is now the number of sheet-coil or sheet-solvent boundaries, and does not yield simply the number of sheet segments. Thus, the lengths of aggregates and the lengths of sheet segments are no longer well-defined for the strip models. These quantities could be defined with a more sophisticated description of aggregates on the strip lattice, for example, by introducing more parameters. For now we try to use a minimum number of parameters and focus on the number of aggregates and the number of sheet proteins in aggregates implied by Eqs.~(\ref{two_filament}) and (\ref{hamilb}), both of which are experimentally measurable properties. 

In Fig.~\ref{2d_plot} we compare qualitatively the results of the $L_{y}=2$ strip models discussed above for $F=1k_{B}T$ and $3k_{B}T$ inter-filament interactions with those of two non-interacting filaments, i.e., $F=0k_{B}T$. We also plot results from the 1D model for the same model parameters. Fig.~\ref{2d_plot}(a) and (b) shows number of aggregates, $\langle \gamma \rangle $, vs. protein coverage for cases A and B, respectively, with $n_{c}=2$. In Fig.~\ref{2d_plot}(a), as $\phi$ increases, the number of aggregates increases from zero and reaches a maximum, then decreases toward zero at maximum protein coverage. Case A yields the results of the 1D model when $F=0k_{B}T$. Overall, increasing $F$ rapidly suppresses the number of aggregates. In case B, the location of the maximum number of aggregates occurs at higher protein coverage when compared with the 1D model when $F=0k_{B}T$. Also, increasing $F$ seems to decrease the numbers of aggregates more slowly for case B when compared with case A for the same model parameters. There are also fewer aggregates in case B when compared to case A. 

The number of sheet proteins in filaments, $\langle \theta \rangle$, is plotted in Fig.~\ref{2d_plot} (c) and (d) for cases A and B, respectively. As protein coverage increases the number of sheet proteins in aggregates increases, more rapidly for increasing $F$. Both models A and B yield essentially the same results for the number of sheet proteins in aggregates for non-zero cases of $F$. When $F=0k_{B}T$, model A predict more sheet proteins in aggregates at low protein coverage when compared to model B, while at high protein coverage model B contains more sheet proteins in aggregates than model A. Thus, overall increasing interchain interaction, $F$, seems to increase the numbers of sheet proteins, but also seems to decrease the numbers of aggregates. This means the number of sheet proteins in aggregates increases rapidly with $F$, a fact consistent with increasing sheet content. This must mean that the size of aggregates and the length of sheet segments increase with $F$.

\section{Comparison to Experiment}
\label{experimental}
Of course, the most important test of a model is whether it can yield results in agreement with experimental observations. In this section, we compare model predictions with the experimental results on A$\beta$(1-40) in Ref.~\onlinecite{terzi} and on Curli fibrils in Ref.~\onlinecite{hammer}. In their work, Terzi et al.~\cite{terzi} used CD spectroscopy, titration calorimetry, and analytical centrifugation to analyze the self-association of A$\beta$(1-40). In aqueous solutions, they showed that A$\beta$(1-40) exhibited a reversible, concentration-dependent sheet-coil transition. Using CD spectroscopy, they obtained the fraction of sheet proteins in aggregates, taken at different concentrations. For our purposes, since the stable oligomer of A$\beta$(1-40) could be the dimer~\cite{selkoe}, we use $n_{c}=2$ in Eq.~(\ref{hamilb}) and calculate Eq.~(\ref{therm4}). We also tried to use the $n_{c}=2$ 1D model described by Eq.~(\ref{latticegas}), which did not produce an acceptable fit. The strip model does produce a good fit for $A\beta$(1-40) aggregates, and is consistent with experimental results.~\cite{selkoe, teplow}

\begin{figure}
	\begin{center}
	\includegraphics[width = \columnwidth]{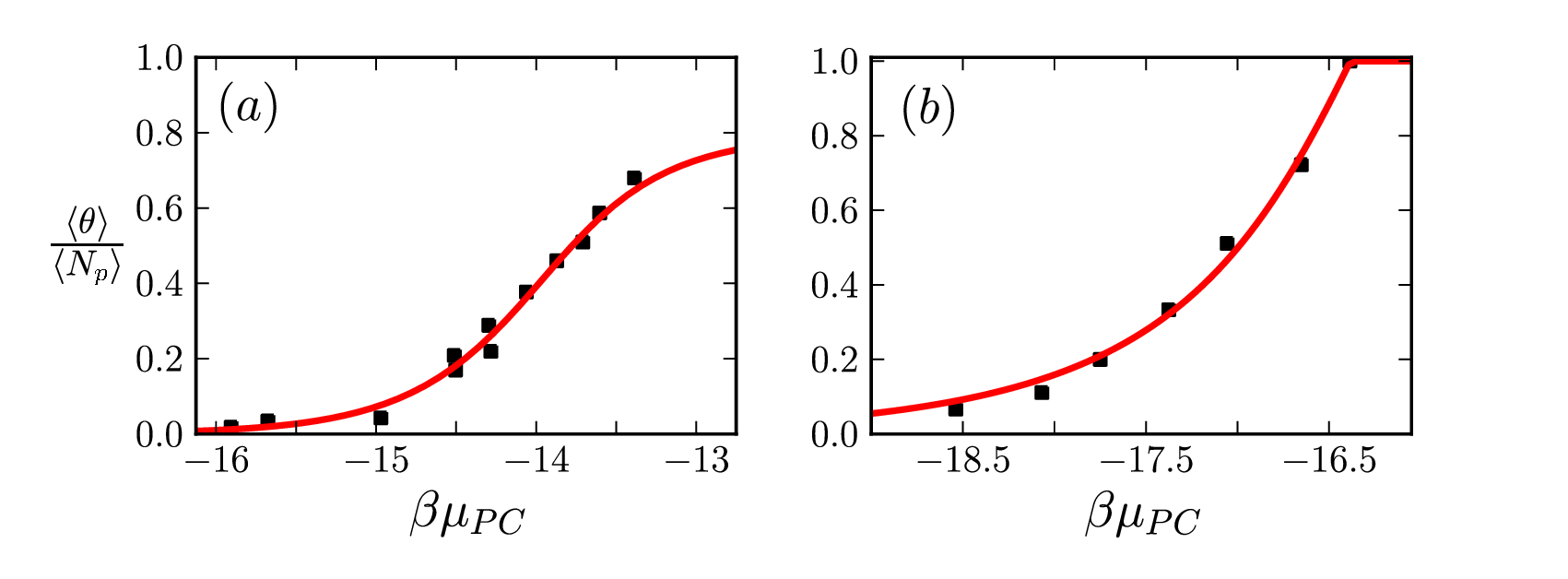}
	\caption{(Color online) \textcolor{black}{ In (a), the fraction of sheet proteins in $A\beta$(1-40) aggregates, $\langle \theta \rangle$/$\langle N_p \rangle$, is fitted to the results of the Terzi et al. experiment~\cite{terzi}. In (b), the fraction of sheet proteins in Curli fibrils is fitted to the scaled results of the Hammer et al. experiment~\cite{hammer}. For the Terzi data, fit parameters were $P_1 \approx K \approx A \approx 0$ kcal/mol, $R_1=0.35$ kcal/mol, and $F=16.4$ kcal/mol. For the Hammer data, $P_1=7.26$ kcal/mol, $K=2.2$ kcal/mol, $R_1\approx0$ kcal/mol, and $A=1.2$ kcal/mol. In (a) we used case B of the strip models with $n_{c}=2$ and Eq.~(\ref{therm4}) as the fit function, whereas in (b) we used the 1D model with $n_c = 2$ for aggregation and Eq.~(\ref{therm2}) as the fit function. In both cases $q=2$, and Eq.~(\ref{therm2}) is divided by $\langle N_p \rangle$ for (a) the strip model and (b) the 1D model, respectively.} }
	\label{fits}
	\end{center}
	\end{figure}

\textcolor{black}{To work with experimental concentration, $c$, we must also specify the other chemical potential contributions in Eq.~(\ref{equib}) for $A\beta$(1-40): $\mu_{ST}$, $\mu_{SR}$ and $\mu_{PV}$. We then calculate $\mu_{PC}$ from Eq.~(\ref{equib}) using the experimental concentrations, and then insert $\mu_{PC}$ into Eq.~(\ref{strip_pf}), from which relevant thermodynamical properties are obtainable. Additionally, in our calculations a 1 mM reference was used in computing the contributions to the solution chemical potential from the experimental concentrations. For A$\beta$(1-40), we have $\mu_{ST}+\mu_{SR}\approx -29$ kcal/mol~\cite{ferrone, hill}. In Ref.~\onlinecite{ferrone}, $\mu_{PV}$ for hemoglobin was found to be approximately $0.75*(\mu_{ST}+\mu_{SR})$. We use a similar result for $\mu_{PV}$ for $A\beta$(1-40), but in reality $\mu_{PV}$ could be larger since $A\beta$(1-40) aggregates may be more flexible than hemoglobin aggregates. We substitute Eq.~(\ref{hamilb}) into Eq.~(\ref{strip_pf}), then Eq.~(\ref{strip_pf}) into Eqs.~(\ref{therm1}) and (\ref{therm4}), and normalize both quantities with respect to by $L_y N_T$. Eq.~(\ref{therm1}) divided by Eq.~(\ref{therm4}), $\langle \theta \rangle / \langle N_p \rangle$, the $\beta$-sheet fraction, is used as our fitting function. The results are plotted in Fig.~\ref{fits}(a).} We calculate as a measure of the quality of the fit the quantity $\eta/N_d \equiv \sqrt{ \sum_{k} \left( \langle \theta_{k} \rangle - \theta_{k} \right)^{2}}/N_d$, where $\langle \theta_{k} \rangle$ is the theoretical value at the $kth$ concentration, $\theta_{k}$ is the experimental value, and $N_d$ is the number of data points in the experiment~\cite{terzi}. The fit yields reasonable free energies at room temperature, $P_1 \approx K \approx A \approx 0$ kcal/mol, $R_1=0.35$ kcal/mol, and $F=16.4$ kcal/mol, and overall a good fit with $\eta/N_{d}$=0.007. 

\textcolor{black}{With $n_{c}=2$, the fitted parameters of our model suggest that $A\beta$(1-40) aggregates will grow easily as indicated by $A\approx0$ kcal/mol, and with $F=16.4$ kcal/mol, the proteins in aggregates are strongly favored to be in the sheet state and bonded with a neighbor in the y-direction. With $K\approx0$ kcal/mol, the aggregates are dominated by sheet structure, and very little coil structure. A fitting value of $R_1=0.35$ kcal/mol suggests that the proteins in aggregates must first overcome an energy barrier before converting from the coil state to the sheet state. Aggregates that form propagate in the x-direction, and the propagation is primarily driven by interactions between sheet proteins in the y-direction rather than directly by the interactions along the x-direction as indicated by $P_{1}\approx 0$ kcal/mol. Thus, once nuclei that are dominated by sheet structure form, aggregates will grow in the x-direction.}
 
Hammer, et al.~\cite{hammer} studied fibrils called Curli. These non-branching, $\beta$-rich fibrils are produced by enteric bacteria, such as E. Coli, and are composed of multiple types of proteins. The major subunit is the CsgA protein which is nucleated into fibrils by another protein, CsgB. Since our model contains only identical proteins, we assume no difference between CsgA, and others, in Curli fibrils. We test our model on the experiment carried out by Hammer, et al., where aggregates of different concentrations of CsgB were detected by Thioflavin T, and TEM analysis at various concentrations revealed the ultrastructure of aggregates at the steady state~\cite{hammer, wetzel}. Since the experiments used Thioflaven T, which binds to fibrils~\cite{bianca}, we scale the florescence data with respect to the fluorescence signal of the highest concentration examined ($c_{0}=43 \mu M$ in their experiments). \textcolor{black}{Here we plot the relative $\beta$-sheet content, not the absolute as with the Terzi data, and divide the number of sheet proteins in filaments, $\langle \theta \rangle$, by $\langle \theta \rangle_{0}$, which is the fluorescence signal at $c_{0}$. The 1D model for aggregation produced an acceptable fit, but the size of a critical nucleus for Curli fibrils is not currently known, so we choose $n_{c}=2$ and substitute Eq.~(\ref{latticegas}) into Eq.~(\ref{grand}). Then plugging Eq.~(\ref{grand}) into Eqs.~(\ref{therm1}) and (\ref{therm4}), we use as our fit function $\langle \theta \rangle / \langle N_p \rangle$.} The data points for different concentrations of CsgB and the theoretical fit are plotted in Fig.~\ref{fits}(b). At room temperature, we find for CsgB $\mu_{ST}+\mu_{SR} \approx -32$ kcal/mol and $\mu_{PV}\approx-25$ kcal/mol. The fitting parameters for the Hammer data were $P_1=7.26$ kcal/mol, $K=2.2$ kcal/mol, $R_1\approx0$ kcal/mol, and $A=1.2$ kcal/mol, and overall a good fit with $\eta/N_d$=0.008, where $N_{d}$ is the number of Curli fibril data points.  

\textcolor{black}{For Curli fibrils, the fitting value of $A=1.2$ kcal/mol suggests that nuclei will form after small assemblies overcome an energy barrier. Since $K=2.2$ kcal/mol, proteins tend to form aggregates. Additionally, $P_1=7.26$ kcal/mol provides strong attraction between sheet proteins, thus monomers in the aggregate will preferentially convert to the sheet state over the coil state. With $R_1\approx0$ kcal/mol, sheet proteins aggregate without overcoming an energy barrier and can covert easily from coil. Thus, the model predicts that the transition from CsgB monomers to Curli fibrils is largely determined by interactions between sheet proteins, and the fibrils largely contain $\beta$-structure.}

\section{Three-state Potts Model for Helix-Sheet-Coil Aggregates}
\begin{figure}
	\begin{center}
	\includegraphics[width = \columnwidth]{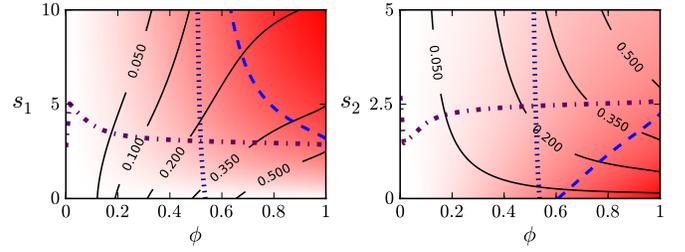}
	\caption{(Color online) (a) Normalized $\langle \theta_1 \rangle$ at a particular ($s_1$, $\phi$) may vary from zero (white color) to one (solid red color). Additionally, contour lines specify the value of $\langle \theta_2 \rangle$ at a particular ($\phi$, $s_{1}$). (b) $\langle \theta_1 \rangle$ and $\langle \theta_2 \rangle$ with the same identifications as in (a) except each quantity is evaluated at a particular ($\phi$, $s_2$). A dotted line indicates equal populations of solvent and proteins in aggregates, a dashed line in both plots indicates sheet-coil/helix transitions, $\langle \theta_{1} \rangle$=0.5, with the remaining proteins either helix or coil, and a black contour line labeled 0.5 in both plots indicates helix-coil/sheet transitions, with $\langle \theta_{2} \rangle$=0.5, and the remaining proteins either sheet or coil. A dashed-dotted line in both plots indicates equal fractions of helix and sheet. Both $\langle \theta_1 \rangle$ and $\langle \theta_2 \rangle$ have been normalized with respect to system size $N_T$. In all cases, $n_{c}=2$, and unless otherwise stated $K=2k_{B}T$, $P_{1}=1k_{B}T$, $P_{2}=1k_{B}T$, $R_{1}=1k_{B}T$, $R_{2}=0.5k_{B}T$, $R_{3}=0k_{B}T$ and $A=1k_{B}T$.}
	\label{3plot}
	\end{center}
	\end{figure}

In this section, we study protein aggregation based on a 3-state (e.g., helix-sheet-coil) 1D lattice-gas model. The lattice-gas Hamiltonian for aggregates containing helix, sheet, or coil conformations is written similarly to Eqs.~(\ref{latticegas}-\ref{nucleus}), except we add interaction terms for helical proteins as given by
\small
\begin{eqnarray}
\label{three_state}
-\beta \mathcal{H}_{fil} &=& \sum_{i=1}^{N_{T}-1} \left\{ P_{1}\delta( t_{i}, 1 ) + P_{2}\delta( t_{i}, 2 ) + K \right\} n_{i} n_{i+1}  \nonumber \\
&-& \sum_{i=1}^{N_{T}-1} R_{1}~\chi(n_{i}, n_{i+1}) \left[ \delta(t_{i},1) n_{i} + \delta( t_{i+1}, 1) n_{i+1} \right] \nonumber \\
&-& \sum_{i=1}^{N_{T}-1} R_{0}~\chi(n_{i}, n_{i+1}) \left[ \delta(t_{i},2) n_{i} + \delta( t_{i+1}, 2) n_{i+1} \right] \nonumber \\ 
&-& \sum_{i=1}^{N_{T}-1} R(t_{i}, t_{i+1}) \chi(t_{i},t_{i+1}) n_{i} n_{i+1} - \beta \mathcal{H}_{ps}^{n_c}   \\
\label{nuke}
-\beta \mathcal{H}_{ps}^{n_{c}} &=& -  \sum_{i=1}^{N_{T}-n_{c}-1} A \chi( n_i, n_{i+n_{c}} ) \prod_{j=i+1}^{i+n_{c}-1} \delta( n_j, 1 ) 
\end{eqnarray}
\normalsize
where $N_T$ is the size of the lattice, and the notation for $R(t_{i}, t_{i+1})$ was discussed in Section~\ref{systems}. After substituting Eq.~(\ref{nuke}) into Eq.~(\ref{three_state}), then plugging into Eq.~(\ref{grand}), we may define relevant thermodynamical quantities as 
\begin{eqnarray}
\label{therm7}
\langle \nu_{j} \rangle &\equiv& \frac{1}{2} \frac{\partial}{\partial R_{j}} \ln \mathcal{Q} \\ 
\label{therm6}
\langle \theta_{i} \rangle &\equiv& \frac{\partial}{\partial P_{i}} \ln \mathcal{Q} 
\end{eqnarray}
where  $\langle \theta_{i} \rangle$ refers to the fraction of sheet, $i=1$, or helix, $i=2$, and $j=0,1,2$ in $\langle \nu_j \rangle$ refers to helix-coil/or solvent, sheet-coil/or solvent, or sheet-helix interfaces, respectively. With these definitions, the number of helix segments is $(\langle \nu_0 \rangle + \langle \nu_2 \rangle)/2$, and the number of  sheet segments is $(\langle \nu_1 \rangle + \langle \nu_2 \rangle)/2$. In Fig.~\ref{3plot} we imposed periodic boundary conditions and computed phase plots for these quantities in the thermodynamic limit. In general, the 3-state model yields richer behaviors than the 2-state model, because helical proteins may also participate in the binding of aggregates. We plot the number of sheet proteins in filaments, $\langle \theta_{1} \rangle$, and the number of helical proteins in filaments, $\langle \theta_{2} \rangle$ vs. $\phi$ and $s_{1}$ in Fig.~\ref{3plot}(a), and vs. $\phi$ and $s_{2}$ in Fig.~\ref{3plot}(b). Additionally, $s_i\equiv\exp(P_i)$ for $i=1,~2$ and the protein coverage $\phi$ is given by Eq.~(\ref{con}). 
	

In Fig.~\ref{3plot}(a), the locations of equal parts helix and sheet proteins in filaments at medium to high protein coverage occurs when $s_1 \approx s_2$, that is the sheet and coil interaction energies are roughly the same magnitude. As $\phi$ decreases the helix/sheet curve occurs for $s_1 > s_2$ with $s_1$ slowly increasing. The sheet-coil/helix transition location, $\langle \theta_1 \rangle =0.5$, is only weakly dependent on large values of $s_1$, but once the sheet interactions weaken and become close in magnitude to helical interactions, the transition locations tend to higher protein coverage, where eventually $s_1\approx s_2$. A transition to majority helical proteins in aggregates occurs only when sheet protein interactions are weaker than attractive, helical protein interactions, that is $s_1<s_2$ with $s_2>1$. Additionally, the number of sheet proteins in filaments, $\langle \theta_{1} \rangle$, is maximal at high values of $s_{1}$ and large protein coverage, which decreases in every direction from this region. Meanwhile, the number of helical proteins in filaments, $\langle \theta_{2} \rangle$, is maximal at low values of $s_1$, and high protein coverage, and decreases in every direction from this region. 

In Fig.~\ref{3plot}(b), the locations of equal parts helix and sheet proteins is mainly independent of $\phi$ and occurs at $s_{2}\approx2.5$ for high $\phi$, that is $s_1\approx s_2$. As $\phi$ decreases, the locations of helix/sheet transitions occur when $s_1 > s_2$ with $s_2$ slowly decreasing. The transition to majority helical proteins in aggregates occurs for $s_2>s_1$, with the locations of transitions occurring at smaller protein coverage as $s_2$ increases. Sheet-coil/helix transitions occur at progressively higher protein coverage for increasing $s_2$ and disappear when $s_2\gtrsim2.5$, that is once helical interactions become stronger than sheet interactions. Like in Fig.~\ref{3plot}(a), when the sheet and helical interactions are attractive, sheet proteins in aggregates dominate at high protein coverage when $s_1> s_2$, and helical proteins dominate at high $\phi$ when $s_1<s_2$.

\section{Discussion and Conclusions}
We have found for the 2-state models with attractive interactions between sheet proteins two regimes: small, largely unstructured aggregates at low protein concentrations, and long sheet dominated filaments at high protein concentrations. The transition from one regime to the other is largely concentration driven, but with the inclusion of nuclei at low concentrations, we found in Fig.~\ref{allplot}(a-b) and (e-f) that fewer filaments form as the size of the nuclei increases. At high concentration, the number of proteins in filaments, and those in the filaments that are sheet, are largely independent of $n_{c}$. We also proposed in addition to the 1D model for aggregation, a quasi-1D model that more realistically captures the nucleation process, where the nuclei structure contains at least two layers of protein that is perpendicular to the direction of propagation of the aggregate, thus the nuclei is a quasi-1D structure. We found that when the interactions between different layers was strongly attractive, the quasi-1D model yielded essentially the same results as the 1D model for the number of sheet proteins in aggregates. When using the same fit parameters, the number of aggregates showed a strong dependence on $F$, where increasing $F$ suppressed the number of aggregates in both strip models, but more significantly for two-filament model when compared to the quasi-1D nuclei model.


We tested the predictions with the 2-state (coil-sheet) 1D model, where the fraction of sheet proteins in aggregates, $\langle \theta \rangle / \langle N_p \rangle$, was used to compare to the experimental results of A$\beta$(1-40) using the strip model for fibrils, and the results of Curli fibrils using the 1D model for fibrils. Fits of both data sets yielded very good agreement. Each of these proteins aggregate into amyloid fibrils through potentially different pathways, thus our model could potentially be applied to a wide variety of pathways in which amyloid fibrils are formed at different concentrations.

For the 3-state model, we found transitions between three regions: sheet dominated regions when helical conformation interactions are weak, helical dominated regions when sheet conformation interactions are weak, and coil aggregates dominate when helical and sheet conformation interactions are weak. In reality, for protein fibrils only the first of the three cases is experimentally relevant. Our model results primarily differ from those of the recent WSME model for aggregation~\cite{zamparo}, which is a peptide bond based model, since it does not consider interactions between helix and coil proteins, only interactions between sheet proteins. \textcolor{black}{By using Potts models in a grand canonical ensemble, our approach to aggregation is quite general and could allow the possibility for helix and coil proteins to participate in aggregation.} The Potts model has the advantage over other simpler models for aggregation because it allows for more conformational states to be considered for proteins, a feature which may prove useful as future experiments involving these characteristics become accessible. 

In conclusion, we have developed statistical mechanical approaches to describe the aggregation of proteins into fibrils in equilibrium. Protein folding and aggregation involve a large number of degrees of freedom, thus it is important to make simplifications when possible. The 1D and quasi-1D statistical mechanical models proposed here have a few parameters and are exactly solvable. For some peptides responsible for neurodegenerative diseases, such as A$\beta$, it is not yet clear whether small oligomers and nuclei are thermodynamically stable, but here we assumed that assemblies from nuclei to fibrils are thermodynamically stable. Calculated thermodynamic quantities mimic certain measurable properties of amyloid fibrils, such as the number of aggregates, the number of sheet segments, and the average lengths of filaments and sheet segments. In order to further test our models, experiments such as AFM measurements of fibril lengths, as was done by van Raaij~\cite{dutch_2008} et al., CD spectra of the sheet content at different concentrations, like in the Terzi data~\cite{terzi}, and also the ThT experiments as in the work of Hammer et al.~\cite{hammer}, should be carried out for various protein species. Additionally, proteins that are known to exhibit more than just 2-state folding ought to be further studied. The 3-state model presented here has the power to capture a more complicated aggregation phenomena where conformations such as helix (and others) may play a role when protein monomers join larger aggregates. With more experimental data, we will be able to draw effectively quantitative comparisons between proteins that aggregate and compile a table of parameters based on our model.

\section*{Acknowledgments}
We would like to thank J. van Gestel and acknowledge F. Ferrone for useful discussions. 


\end{document}